\documentstyle[pra,aps,amssymb,epsf]{revtex}

\begin{document}


\title{Some methods for solution of quantum detection and
measurement problems\thanks{\bf Translation of the original Russian
paper: {\it B. A. Grishanin, Tekhnicheskaya Kibernetika, {\bf 11} (5),
pp. 127--137 (1973)}, \bf submitted 18 January 1972.}}
\author{B.\ A.\ Grishanin\thanks{grishan@comsim1.phys.msu.su}}
\address{International Laser Center and Department of Physics\\
M.\ V.\ Lomonosov Moscow State University, 119899 Moscow, Russia}
\date{January 29, 2003}
\maketitle

\thispagestyle{empty}

\section{Introduction}
\label{sec:intro} The problem of detection and measurement optimization
at the output of a quantum communication channel, under its modern
interpretation \cite{Helstrom}, can be described in the following
manner.

Let a quantum channel $\{H,\hat\rho(\lambda)\}$ is given. Here $H$ is
the Hilbert space where the operators are defined, which represent (in
accordance with the basic principles of quantum theory) the quantum
observables of the channel output; $\hat\rho(\lambda)$ is the density
matrix, which represents the quantum state of the output observables
and depends on input classical observables $\lambda\in\Lambda$. The
same way as in the classical decision theory \cite{bolshakov} a set
$\tilde\Lambda$ of the possible decisions $\tilde\lambda$, which are
represented now by a final result of some quantum measurement procedure
$\gamma$. The problem is to find an optimal procedure $\gamma$ in the
standard sense of the decision theory. The latter means that the
optimization quality is represented by a conditional risk function
\begin{equation}\label{rg}
r(\gamma|\lambda)=\int c(\lambda,\tilde\lambda)\,{\cal
P}_\gamma(d\tilde \lambda|\lambda),
\end{equation}

\noindent where $c(\lambda,\tilde\lambda)$ is the ``cost'' (pay)
function and ${\cal P}_\gamma(d\tilde \lambda|\lambda)$ is the
conditional probability distribution depending on the procedure
$\gamma$ and the conditional channel output state $\hat\rho(\lambda)$.

We suggest the following representation (modified with respect to
\cite{Helstrom}) of the measurement procedure. In accordance with
the axioms of quantum theory, a direct quantum measurement of any
physical observables represented with a commuting set of operators
$\hat\Theta$, which are defined in the tensor product $H\otimes
H'$ of the given space $H$ and any additional auxiliary space
$H'$. The latter introduces some channel-independent quantum
system with an arbitrary density matrix $\hat\rho\,'$, so that the
joint density matrix of the output and the auxiliary subsystems
takes the form $\hat\rho(\lambda) \otimes\hat\rho\,'$. We call
this indirect measurement. Directly measured variables
$\hat\Theta$ yield spectral values $\tilde\lambda\in
\tilde\Lambda$, which represents the decisions of interest. In
accordance with the quantum measurement postulates, we get
\begin{equation}\label{Pg}
{\cal P}_\gamma(d\tilde\lambda|\lambda)={\rm
Tr}\,[\hat\rho(\lambda) \otimes\hat\rho\,']\hat E(d\tilde\lambda),
\end{equation}

\noindent where $\hat E$ is an orthogonal decomposition of unit
\cite{Akhieser} for operators
\begin{displaymath}
\hat\Theta=\int\limits_{\tilde\Lambda}\tilde\lambda\, \hat
E(d\tilde\lambda).
\end{displaymath}

Performing in Eq. (\ref{Pg}) an averaging over the auxiliary
system, we get
\begin{equation}\label{PgE}
{\cal P}_\gamma(d\tilde\lambda|\lambda)={\rm
Tr}\,\hat\rho(\lambda)\hat{\cal E}(d\tilde\lambda),
\end{equation}

\noindent where
\begin{equation}\label{Edl}
\hat{\cal E}(d\tilde\lambda)={\rm Tr}_{H'}\hat
E(d\tilde\lambda)\,\hat\rho\,'
\end{equation}

\noindent is the quantum decision function which determines
statistical properties of the measurement results. It is the
quantum analogue of the randomized decision function, that is
conditional probability measure $\mu(d\tilde\lambda|y)$, where $y$
are the classical output observables. Here Eq. (\ref{PgE}) is
analogous to the classical formula ${\cal
P}(d\tilde\lambda|\lambda)=\int\mu(d\tilde\lambda|y) {\cal
P}(dy|\lambda)$. In contrast to $\mu$, Eq. (\ref{Edl}) defines an
operator self-adjoint measure (non-orthogonal, or generalized,
decomposition of unit \cite{Akhieser}) obeying the positivity and
normalization conditions:
\begin{equation}\label{cond}
\hat{\cal E}(\cdot)\geqq 0,\quad \int\hat E(d\tilde\lambda)=\hat I.
\end{equation}

Thus, optimization over the possible setting $(H',\hat\rho\,',\hat
\Theta)$ is reduced to optimization over $\hat{\cal E}$. A similar
approach to the problems of quantum mathematical statistics is
developed in \cite{Holevo}, which appeared after the submission of
this paper; the main results of this paper had been presented in
\cite{Gorkii}.

Here only the Bayes type of optimization problems is under study, where
an a'priory probability distribution ${\cal P}(d\lambda)$ is known and
average risk takes the form
\begin{displaymath}
{\cal R}=\int r(\gamma|\lambda)\,{\cal P}(d\lambda).
\end{displaymath}

\noindent Substituting here expressions (\ref{rg}), (\ref{PgE}) and
introducing operators of the quantum a'posteriori risk
\begin{equation}\label{Rl}
\hat R(\tilde\lambda)=\int c(\lambda,\tilde\lambda)\,\hat\rho(\lambda)
\,{\cal P}(\lambda)
\end{equation}

\noindent (which is an analogue of the product $R(\tilde\lambda|y)
{\cal P}(dy)$ of the classical a'posteriori risk and the output
observables $y$ probability distribution), we arrive at the
minimization problem
\begin{equation}\label{min}
{\cal R}(\hat{\cal E})={\rm Tr}\int\hat R(\tilde\lambda)\, \hat{\cal
E}(d\tilde\lambda)=\min\limits_{\hat{\cal E}},
\end{equation}

\noindent the solution of which is the optimal decision function
$\hat{\cal E}_0$.

In this paper the analysis of the possible methods to get the solution
of the problem (\ref{min}) is presented. For gaussian channels, for the
cases of so called ``simple'' and quadratic cost functions the optimal
decision functions $\hat{\cal E}_0$ and the corresponding optimal
measurement procedures are obtained. The most attention is paid to the
gaussian density matrices
\begin{equation}\label{rho}
\hat\rho(\lambda)=\exp[\Gamma-(\hat x-\lambda)^T Q (\hat x-\lambda)],
\end{equation}

\noindent where $\hat x^T=(\hat x_1,\dots,\hat x_s)$ is a given
vector of operators in the $H$ space with a non-degenerate (only
for simplicity of discussion) $c$-number commutation matrix:
\begin{equation}\label{Cij}
[\hat x,\hat x^T]=(\hat x_i\hat x_j-\hat x_j\hat x_i)=C\hat I, \quad
(i,j=1,\dots,s);
\end{equation}

\noindent $Q>0$ --- a given definitely positive symmetric matrix;
\begin{equation}\label{G}
\Gamma=\ln{\det}^{1/2}|2\sinh\bigl(\sqrt{Q}\,C\sqrt{Q}\bigr)|
\end{equation}

\noindent is the normalization constant. On the base of Eqs.
(\ref{rho}), (\ref{Cij}) the correlation matrix can be calculated as
\begin{equation}\label{K}
K=\Bigl(\frac{1}{2}{\rm Tr}\,[(\hat x_i-\lambda_i)](\hat
x_j-\lambda_j) +(\hat x_j-\lambda_j)(\hat
x_i-\lambda_i)]\hat\rho(\lambda)\Bigr)= \frac{1}{2}C\coth QC.
\end{equation}

Gaussian density matrices arise at the investigation of the
thermodynamically equilibrium linear physical systems with a
Hamiltonian ${\cal H}(\lambda)$ (that is a total energy
represented in terms of coordinates and momenta), which is
quadratic with coordinates and momenta, i.e. corresponds to a
collection of linear oscillators. Typical examples are given by
vibrational circuits, electromagnetic fields in lines, waveguides
and free space. Parameters $\lambda$ in (\ref{rho}) for the most
typical examples correspond to a non-quantum displacement, due to
the presence of a classical signal at their input. Gibbs
distribution \cite{Klauder} $\hat\rho=\exp(\Gamma-\hat{\cal H}/
kT)$ for such systems yields density matrix of the form
(\ref{rho}).

\section{Method of the shifted a'posteriori risk}
\label{section:displaced} If the family $R$ of the a'posteriori risk
operators is representable in the form
\begin{equation}\label{Rls}
\hat R(\tilde\lambda)=\hat{\tilde R}(\tilde\lambda)+\hat A,
\end{equation}

\noindent where operator $\hat A$ does not depend on $\tilde\lambda$,
then the minimization problem is equivalent to minimization
\begin{equation}\label{minnew}
\int\,{\rm Tr}\,\hat{\tilde R}(\tilde\lambda)\,\hat{\cal E}
(d\tilde\lambda) =\min\limits_{\hat{\cal E}}
\end{equation}

\noindent for the same constraints on $\hat{\cal E}$. Such replacement
of the minimization problems takes place when the cost function is
represented in the form
\begin{equation}\label{cc}
c(\lambda,\tilde\lambda)=\tilde{c}(\lambda,\tilde\lambda)+c_0(\lambda).
\end{equation}

\noindent Then the corresponding operators $\hat{\tilde R}(\tilde
\lambda)$, $\hat A$ can be expressed via $\tilde{c},c_0$ with a help of
Eq. (\ref{Rl}).

If the ``shifted'' family $\tilde{R}$ is commutative, then all
operators $\hat{\tilde R}(\tilde\lambda)$ can be represented as
functions of some commutative set of operators $\hat y$. Then the
solution of the problem (\ref{minnew}) is easy to obtain in the form
\begin{equation}\label{E0}
\hat{\cal E}_0(d\tilde\lambda)=\delta(d\tilde\lambda,
\tilde\lambda_0(\hat{y})),
\end{equation}

\noindent where the scalar function is the indicator function of the
subset $e$:
\begin{equation}\label{d}
\delta(e,x)=\left\{\begin{array}{ll}
  0, & x\,\overline{\in}\,e, \\
  1, & x\in e;
\end{array}\right.
\end{equation}

\noindent $\tilde\lambda_0(\hat{y})$ --- scalar function obtained
after minimization with $\tilde\lambda$ the matrix elements
$\tilde{R}_{yy}$ of the operators $\hat{\tilde R}(\tilde\lambda)$
at the $y$-representation. The corresponding measurement procedure
is given then by direct quantum measurement of the variables
$\tilde\lambda_0(\hat{y})$, as far as Eq. (\ref{E0}) is an
orthogonal expansion of unit for the $\tilde\lambda_0(\hat{y})$
operators. As far as $\tilde\lambda_0(\hat{y})$ are commutative,
the decision function (\ref{E0}) can be interpreted as a singular
probability distribution of the classical type
$\mu(d\tilde\lambda|\tilde\lambda_0)=
\delta(d\tilde\lambda,\tilde\lambda_0)$ and the corresponding
measurement as a non-randomized one. Keeping in mind that any
orthogonal expansion may be written in the form of Eq.~(\ref{E0}),
we see that any direct measurement is a quantum analogue of a
non-randomized one. Conversely, the indirect measurement can be
called randomized. In the quantum case, in contrast to classical
Bayes problems, randomization is always necessary when the useful
information is associated with non-commuting observables and their
direct measurement is impossible.

It is easy to prove, that in the problem of choosing of one of the
two possible decisions $\tilde\lambda=\tilde\lambda_1$ or
$\tilde\lambda=\tilde \lambda_2$ it is ever possible to find
operator $\hat A$ to fit the demand of commutativity of the
shifted risk operators $\hat{\tilde R}(\tilde\lambda)$. Hence, in
these kind of problems the optimum is ever found on the
non-randomized --- direct measurements \cite{Helstrom}. In all
other cases, for non-commutative $\hat\rho(\lambda)$, it is the
most probable that the above method is incapable to provide the
commutative shifted risk operators $\hat{\tilde
R}(\tilde\lambda)$. Nevertheless, as it will be shown at Sec. 4,
this method may be useful in combination with the other methods,
which may happen effective in the transformed problem
(\ref{minnew}).

\section{Method of a lower boundary}\label{section:lower}
The general sketch of this method is as follows. The minimal
average risk ${\cal R}(\hat{\cal E})$ is replaced by some lower
boundary ${\cal R}_0(\hat{\cal E})\leqq{\cal R}(\hat{\cal E})$.
For ${\cal R}_0(\hat{\cal E})$ a class $\Sigma$ of the decision
functions $\hat{\cal E}$, which minimize ${\cal R}_0(\hat{\cal
E})$, is determined:
$$
{\cal R}_0(\hat{\cal E})=\min\,{\cal R}_0(\hat{\cal E}).
$$

\noindent If this class $\Sigma$ contains at least one decision
function $\hat{\cal E}_0$ which fits the condition
$$
{\cal R}(\hat{\cal E}_0)={\cal R}_0(\hat{\cal E}_0),
$$

\noindent then $\hat{\cal E}_0$ is the solution of the minimization
problem (\ref{min}), as ${\cal R}(\hat{\cal E}_0)\leqq{\cal R}
(\hat{\cal E})$ for all $\hat{\cal E}$.

One of the simplest lower boundaries is
\begin{equation}\label{lower}
{\cal R}_0(\hat{\cal E})=\int\,r(\tilde\lambda){\rm Tr}\,\hat{\cal
E}(\tilde\lambda),
\end{equation}

\noindent where $r(\tilde\lambda)$ are the minimal eigen values of the
risks $\hat R(\tilde\lambda)$. For the corresponding eigen projectors
we have
\begin{equation}\label{P}
\hat R(\tilde\lambda)\hat P(\tilde\lambda)=r(\tilde\lambda)\hat
P(\tilde\lambda).
\end{equation}

\noindent At deduction of Eq. (\ref{lower}) the inequality ${\rm
Tr}\, [\hat R(\tilde\lambda)-r(\tilde\lambda)\hat I]\hat{\cal
E}(\tilde \lambda)\geqq0$ is used. For example, if all operators
$\hat R(\tilde\lambda)$ are unitary equivalent all the
$r(\tilde\lambda)=r_0$ are equal. Then with use of Eq.
(\ref{lower}) for a finite-dimension space $H$ we get an optimal
operator measure $\hat{\cal E}_0$ in the form of
\begin{equation}\label{E00}
\hat{\cal E}_0(d\tilde\lambda)=\hat
P(\tilde\lambda)\nu(d\tilde\lambda),
\end{equation}

\noindent if only the positive $c$-number measure $\nu$ fits the
normalization condition given by the second relation in Eqs.
(\ref{cond}). In the case of infinite-dimension space $H$ the risk
would be infinite, so this case is of no interest. To the latter
situation the below discussed problem is simplified.

Let there is a Bayes problem with the simple cost function $c(\lambda,
\tilde\lambda)=-\delta(\lambda-\tilde\lambda)$ ($\Lambda=\tilde
\Lambda$ are $s$-dimensional Euclid spaces), gaussian family $\hat
\rho(\lambda)$ described by Eqs. (\ref{rho})--(\ref{K}) and infinitely
wide a'priory distribution $p(\lambda)=d{\cal P}/d\lambda$\,.
Corresponding a'posteriori risk operators (\ref{Rl}) are of the form
\begin{equation}\label{Rll}
\hat{R}(\tilde\lambda)=-p(\tilde\lambda)\hat\rho(\tilde\lambda).
\end{equation}

\noindent Let us show that in this problem the optimal measure
$\hat{\cal E}_0$ is given in the form (\ref{E00}) with
\begin{equation}\label{nu}
\nu(d\tilde\lambda)=\frac{d\tilde\lambda}{{\det}^{1/2}|2\pi C|}\,,
\end{equation}

\noindent and $\hat P(\tilde\lambda)$ are given by the eigen
projectors corresponding to to the maximal eigen value $\mu_0$ of
the density matrices $\hat\rho(\lambda)$:
\begin{equation}\label{rhoP}
\hat\rho(\tilde\lambda)\hat P(\tilde\lambda)=\mu_0\hat
P(\tilde\lambda).
\end{equation}

Now let us divide the space $\tilde\Lambda=\Lambda\ni\lambda$ into
regions $\Lambda_i$, which on the one hand have the dimension
small enough to neglect the dependence of $p(\lambda)$ on
$\lambda$ and on the other hand unlimitedly expand with expansion
of the $p(\lambda)$ function. Then on account of Eq. (\ref{Rll})
we can represent risk (\ref{min}) in the form
\begin{equation}\label{Rnew}
{\cal R}(\hat{\cal E})=-\sum\limits_i
p(\lambda_i)\int\limits_{\Lambda_i}{\rm Tr}\,\hat\rho(\lambda)
\hat{\cal E}(d\lambda).
\end{equation}

\noindent For fixed $i$ a'posteriori risks $R_i(\lambda)=-
p(\lambda_i)\hat\rho(\lambda)$ in the right side of Eq.
(\ref{Rnew}) are now unitary equivalent. For the subsequent
analysis the following asymptotic property is essential for the
family of operators $\hat \rho(\lambda)$ with
$\lambda\in\tilde{\Lambda}_i=\Lambda_i\setminus\delta\Lambda_i$,
that is $\lambda$ belongs to the internal region $\tilde\Lambda_i$
got by subtraction of a small boundary region $\delta\Lambda_i$,
the latter unlimitedly expanding with $p(\lambda)$ but having a
zero relative volume with respect to the one of $\Lambda_i$. For
any in advance given precision these families may be supposed to
act non-trivially only in the subspaces
$\Phi_i(\varepsilon)\subset H$, $\varepsilon\to0$, beyond which,
due to relation $\|\hat\rho(\lambda)\psi\|\leqq\varepsilon
\|\psi\|$, it is acceptable to set $\hat\rho(\lambda)\psi=0$. That
is for unlimited expansion of the regions $\Lambda_i$,
$\delta\Lambda_i$, $\tilde\Lambda_i$ the subspaces $\Phi_i$ are
asymptotically orthogonal to each other with any unlimitedly high
precision given by $\varepsilon$. As it is shown in Appendix A,
for the discussed family $\hat\rho(\lambda)$ the orthoprojectors
$\hat P_i$ onto the introduced above subspaces $\Phi_i$ can be
represented in the form
\begin{equation}\label{PD}
\hat P_i=\int\limits_{\tilde\Lambda_i}\hat P(\lambda)
\frac{d\lambda}{{\det}^{1/2}|2\pi C|}\,,
\end{equation}

\noindent and their dimensionality is $N_i={\rm Tr}\,\hat P_i=
{\det}^{-1/2}|2\pi C|V(\tilde\Lambda_i)$ ($V$ --- the
corresponding phase subspace volume).

Taking into account equation $\hat\rho(\lambda)=\hat P_i
\hat\rho(\lambda)\hat P_i$, arising from the above discussion for
$\lambda\in\tilde\Lambda_i$, and neglecting small boundary regions
$\delta\Lambda_i$, the terms of the sum (\ref{Rnew}) can be
rewritten as
\begin{equation}\label{Ri}
{\cal R}_i(\hat{\cal E}_i)=-p(\lambda_i)
\int\limits_{\tilde\Lambda_i} {\rm Tr}\,\hat\rho(\lambda)\hat{\cal
E}_i(d\lambda),
\end{equation}

\noindent where $\hat{\cal E}_i$ are the measure $\hat{\cal E}$
projection onto $\Phi_i$ fitting the normalization condition
\begin{equation}\label{EN}
\int\limits_{\tilde\Lambda_i}\hat{\cal E}_i(d\lambda)=\hat P_i.
\end{equation}

\noindent Due to the asymptotic orthogonality of $\Phi_i$,
measures $\hat{\cal E}_i$ are independent decompositions of
operators $\hat P_i$ in $\Phi_i$, so minimization of
Eq.~(\ref{Rnew}) over $\hat{\cal E}$ is equivalent to minimization
of Eq.~(\ref{Ri}) over $\hat{\cal E}_i$.

To perform this minimization let us make use of solution (\ref{E00})
with measure $\nu$ of the form (\ref{nu}) taking into account that in
accordance with (\ref{PD}) it fits the normalization condition
(\ref{EN}). On additional account of independence of measure (\ref{nu})
on $i$ this solution (\ref{E00}) defines the optimal measure $\hat{\cal
E}_0$ in the total space of $\tilde\lambda$ values, so that
asymptotically $\hat{\cal E}_0\to\bigoplus_i\hat{\cal E}_i$.

Taking into account Eqs. (\ref{Rll}) and (\ref{rhoP}) we get from Eq.
(\ref{P})
\begin{equation}\label{rp}
r(\lambda)=-p(\lambda)\mu_0,
\end{equation}

\noindent and for the average risk from (\ref{lower}) or (\ref{Rll}),
\begin{equation}\label{RE0}
{\cal R}(\hat{\cal E}_0)=-\int p(\lambda)\mu_0
\frac{d\tilde\lambda}{{\det}^{1/2}|2\pi C|}=-
\frac{\mu_0}{{\det}^{1/2}|2\pi C|}\,.
\end{equation}

\noindent Applying then the zero-temperature method presented in
Appendix A it is easy to calculate
\begin{equation}\label{mu0}
\mu_0=\exp\Bigl\{\frac{1}{2}\,{\rm Tr\Bigl[\ln|2\sinh QC|+ |QC|
\Bigr]}\Bigr\}.
\end{equation}

\noindent Correspondingly, from Eq. (\ref{RE0}) and  (\ref{mu0}) for
$C\to0$ we get the classical limit ${\cal R}=-{\det}^{1/2}(Q/\pi)$.

\section{Method of reducing of the a'posteriori risk operators to a
family of unitary equivalent operators}\label{section:reducing} In some
problems the a'posteriori risks operators may be not unitary
equivalent, as it was in Sec. \ref{section:lower}, but with use of
replacement (\ref{Rls}) it is possible to get such a representation for
the displaced average risk $\tilde{\cal R}(\hat{\cal E})={\cal R}
(\hat{\cal E})-{\rm Tr}\,\hat A$, that the method of a lower boundary
would be applicable.

Let us discuss the case of quadratic cost function
\begin{equation}\label{cll}
c(\lambda,\tilde\lambda)=(\tilde\lambda-\lambda)^T(\tilde\lambda-\lambda)
\end{equation}

\noindent (it is no demand to introduce a more general quadratic form,
as it is always can be reduced to this one). For this case
\begin{equation}\label{Rquad}
\hat
R(\tilde\lambda)=\int(\lambda-\tilde\lambda)^T(\lambda-\tilde\lambda)
\hat\rho(\lambda){\cal P}(d\lambda).
\end{equation}

\noindent Let us denote
\begin{equation}\label{tdr}
\hat{\tilde\rho}=\int\hat\rho(\lambda){\cal P}(d\lambda)
\end{equation}

\noindent and find such operators $\hat u$, that Eq. (\ref{Rquad}) is
reduced to
\begin{equation}\label{Rred}
\hat R(\tilde\lambda)=\hat{\tilde\rho}^{1/2}(\hat u-\tilde\lambda)^T
(\hat u-\tilde\lambda)\hat{\tilde\rho}^{1/2}+\hat A,
\end{equation}

\noindent where $\hat A$ does not depend on $\tilde\lambda$. It is
easy to confirm that these operators are the operators of the
a'posteriori mathematical expectation of variables $\lambda$:
\begin{equation}\label{ua}
\hat u=\int\lambda\hat F(d\lambda),
\end{equation}

\noindent where $$\hat F(d\lambda)=\hat{\tilde\rho}^{-1/2}
\hat\rho(\lambda)\,\hat{\tilde\rho}^{1/2}{\cal P}(d\lambda)$$

\noindent is a non-orthogonal decomposition of unit having a
meaning of a specific quantum a'posteriori probability
distribution of variables $\lambda\,$. Note, that here introduced
analogues of classical a'posteriori characteristics are not unique
(e.g. see \cite{Stratonovich}) and the adequate form depends on a
specific problem.

Operator $\hat A$ corresponding to operators (\ref{ua}) in Eq.
(\ref{Rred}) is equal to
\begin{equation}\label{Au}
\hat A=\int \lambda^T\lambda\,\hat\rho(\lambda){\cal P}(d\lambda) -
\hat{\tilde\rho}^{1/2}\hat u^T\hat u\hat{\tilde\rho}^{1/2}
\end{equation}

\noindent and the corresponding risk is $${\cal R}_A={\rm Tr}\,\hat A
={\rm Tr}\,(K_\lambda-\tilde K_u),$$

\noindent where $K_\lambda$ is a'priori correlation matrix of
variables $\lambda$, for which the equality ${\bf\rm M}\lambda=0$
is suggested; $\tilde K_u$ denotes the condition-less correlation
matrix of $\hat u$ depending on the condition-less density matrix
(\ref{tdr}).

This way obtained operators of the shifted risk $$\hat{\tilde R}(\tilde
\lambda)=\hat{\tilde\rho}^{1/2}(\hat u-\lambda)^T(\hat u-\lambda)\,
\hat{\tilde\rho}^{1/2}$$

\noindent are not unitary equivalent. Although, if in equation $\tilde
{\cal R}(\hat{\tilde{\cal E}})={\rm Tr}\,\hat{\tilde R}(\tilde\lambda)
\hat{\cal E}(d\tilde\lambda)$ for the average risk we substitute
\begin{equation}\label{tE}
\hat{\tilde\rho}^{1/2}\hat{\cal E}(d\tilde\lambda)
\hat{\tilde\rho}^{1/2}=\hat{\tilde{\cal E}}(d\tilde\lambda),
\end{equation}

\noindent then it reduces to
\begin{equation}\label{tRtE}
\tilde {\cal R}(\hat{\tilde{\cal E}})={\rm Tr}\int (\hat
u-\tilde\lambda)^T(\hat u-\tilde\lambda)\hat{\tilde{\cal E}}
(d\tilde\lambda),
\end{equation}

\noindent where $\hat{\tilde{\cal E}}$ is a new positive measure
normalized the following way
\begin{equation}\label{NtE}
\int\hat{\tilde{\cal E}}=\hat{\tilde\rho}
\end{equation}

\noindent instead of the second equality in Eq. (\ref{cond}). For the
gaussian density matrix (\ref{rho}) and gaussian ${\cal P}(d\lambda)$
we have $$\hat{\tilde\rho}=\exp(\tilde\Gamma-\hat x^T\tilde{Q}\hat x)$$

\noindent with $\tilde Q=(1/2)C^{-1}\coth^{-1}[2(K+K_\lambda)C^{-1}]$.
At this from Eq. (\ref{ua}) we get $$\hat u=V\hat
x=K_\lambda(K+K_\lambda)^{-1}\cosh(C\tilde Q)\hat x.$$

\noindent Making use of operators $\hat u_0=K_\lambda(K+
K_\lambda)^{-1}\hat x$, which determine the optimal operator estimates
with no account of the commutativity condition \cite{Stratonovich}, we
represent this expression in the form $$\hat u=\cosh C_0\tilde
Q_0\,\hat u_0,$$

\noindent where $C_0=K_\lambda(K+ K_\lambda)^{-1}C(K+
K_\lambda)^{-1} K_\lambda$, $\tilde
Q_0=[(K+K_\lambda)^{-1}K_\lambda]^{-1}\tilde Q
[K_\lambda(K+K_\lambda)^{-1}]^{-1}$ are, correspondingly, the
commutator of the variables $\hat u_0$ and the responding to it
$Q$-matrix in the expression
$\hat{\tilde\rho}=\exp(\tilde\Gamma-\hat u_0^T\tilde Q_0\hat u_0)$
for the transformed gaussian density matrix of the output state.

As a result of the presented transformations, for the multiplier
of $\hat{\tilde{\cal E}}$ in Eq. (\ref{tRtE}) we have now unitary
equivalent operators $(\hat u-\lambda)^T(\hat u-\lambda)$, due to
which with use of the lower boundary (\ref{lower}) after replacing
$\hat{\cal E}$ to $\hat{\tilde{\cal E}}$ we immediately get the
solution in the form of Eq. (\ref{E00}) where projectors $\hat
P(\lambda)$ correspond to minimal eigen value $r_0$ of operators
$(\hat u-\lambda)^T(\hat u-\lambda)$:
\begin{equation}\label{uu}
(\hat u-\lambda)^T(\hat u-\lambda)\hat P(\lambda)=r_0\hat P(\lambda).
\end{equation}

\noindent Corresponding measure $\nu$ is to be found from the
normalization condition (\ref{NtE}) which results in equation
\begin{equation}\label{NP}
\int\hat P(\tilde\lambda)\nu(d\tilde\lambda)=\hat{\tilde\rho}.
\end{equation}

If the equation
\begin{equation}\label{KC}
\tilde{K}_u-\frac{1}{2}|C_u|>0
\end{equation}

\noindent holds, then on account of that $\hat P(\lambda)$ is a family
of the gaussian density matrices with the correlation matrix $K_0=(1/2)
|C_u|$ (see Sec. \ref{section:lower}), from Eq. (\ref{NP}) we get
measure $\nu$ in the form of the gaussian probability measure
\begin{equation}\label{nulambda}
\nu(d\tilde{\lambda})={\det}^{-1/2}\Bigl[2\pi\Bigl(\tilde{K}_u-
\frac{1}{2}|C|_u\Bigr)\Bigr]\,\exp\Bigl[-\frac{1}{2}\tilde{\lambda}^T
\Bigl(\tilde{K}_u-\frac{1}{2}|C_u|\Bigr)^{-1}\tilde{\lambda}\Bigr]
d\tilde{\lambda}.
\end{equation}

\noindent Expressing the original measure $\hat{\cal E}_0$ via the
above discussed extreme measure $\hat{\tilde{\cal E}}_0$ in accordance
with Eq.~(\ref{tE}) we get the optimal quantum decision function
\begin{equation}\label{Eopt}
\hat{\cal E}_0(d\tilde{\lambda})=\hat{\tilde{\rho}}^{-1/2}\hat P(\tilde
\lambda)\hat{\tilde{\rho}}^{-1/2}\nu(d\tilde{\lambda}).
\end{equation}

It is capable to interpret this solution in a simpler equivalent form.
For this purpose let us take into account that projectors $\hat
P(\tilde\lambda)$ are of the form $\psi(\tilde\lambda)
\psi^+(\tilde\lambda)$, where $\psi(\tilde\lambda)$ are normalized
``Glauber'' eigen vectors of the operators $(\hat u-\tilde\lambda)^T
(\hat u-\tilde\lambda)$. Then
\begin{equation}\label{tP}
\hat{\tilde{P}}(\tilde\lambda)=\bigl[{\rm Tr}\,\hat{\tilde{\rho}}^{-1}
\hat P(\tilde\lambda)\bigr]^{-1}\hat{\tilde{\rho}}^{-1/2}\hat
P(\tilde\lambda)\hat{\tilde{\rho}}^{-1/2}
\end{equation}

\noindent are also projectors of the form $$\hat P(\tilde\lambda)=
\varphi(\tilde\lambda)\varphi^+(\tilde\lambda)\quad
\Bigl(\varphi=(\psi^+\hat{\tilde\rho}^{-1}\psi)^{-1/2}
\hat{\tilde\rho}^{-1/2}\psi\Bigr).$$

\noindent They are, as it is easy to show after multiplication of
Eq.~(\ref{uu}) both on the left and right by $\hat{\tilde\rho}^{-1/2}$
and applying twice the commutation rule \cite{Stratonovich}
$\hat{\tilde{\rho}}^{-1/2}\hat u=\exp(-C_u\tilde{Q}_u)\,\hat
u\hat{\tilde{\rho}}^{-1/2}$ ($\tilde{Q}_u=V^{-1T}\tilde{Q}V^{-1}$), the
eigen projectors of a new generally non-Hermitian quadratic form:
\begin{equation}\label{qform}
[\exp(-C_uQ_u)\hat u-\tilde\lambda]^T[\exp(-C_uQ_u)\hat u-
\tilde\lambda]\hat{\tilde P}(\tilde\lambda)=r_0\hat{\tilde
P}(\tilde\lambda).
\end{equation}

\noindent Here again, like in Eq. (\ref{uu}), $$r_0=\frac{1}{2}\,{\rm
Tr}\,|C_u|.$$

\noindent From Eq. (\ref{qform}) it is easy to get that asymptotically
with $C_u\to0$ projectors $\hat{\tilde P}(\tilde\lambda)$ and
$\hat{P}(\tilde\lambda)$ coincide. With use of $\hat{\tilde P}(\tilde
\lambda)$ Eq. (\ref{Eopt}) takes the form
\begin{equation}\label{Eoptm}
\hat{\cal E}_0(d\tilde\lambda)=\hat{\tilde P}(\tilde\lambda)
\tilde{\nu}(d\tilde\lambda),
\end{equation}

\noindent where measure $\tilde\nu$ may be calculated also with use of
normalization condition (\ref{cond}) (the second relation). At the
limit $C_u\to0$ Eq. (\ref{Eoptm}) differs from optimal measure
$\hat{\cal E}_0$, calculated in Sec.~\ref{section:lower} for the case
of simple cost function, by the change $\hat x\to\hat u$ and applying
the corresponding definitions for the projectors $\hat P(\tilde
\lambda)$ and commutation matrix $C$. Hence, the responding optimal
measurement procedures are also identical.

If condition (\ref{KC}) does not hold then Eq. (\ref{nulambda}) is not
applicable. Not going into detailed analysis of this case, we simply
mark here, that its qualitative specificity may be revealed by analysis
of the transition from the non-degenerate matrix $\tilde{K}_u-(1/2)
|C_u|$ to the degenerate one. At that, over one or several linear
combinations of variables $\hat u$ measure (\ref{nulambda}) will be
localized at zero (it is the a'priori mathematical expectation). This
means that the corresponding variables do not need refinement on the
base of available information. If matrix $\tilde{K}_u-(1/2) |C_u|$ is
not positive than, from the physical qualitative considerations which
are supplied with corresponding mathematical analysis, it is clear that
the above described degeneration persists.

\section{Method of information constraints}\label{section:info} This
method is based \cite{Mamaev} on reducing, with use of a small
additional term, the linear optimization problem (\ref{min}) and
(\ref{cond}) (the second relation) to an appropriate non-linear one
which lets using the standard differential variation methods. Like in
\cite{Mamaev} it is possible here to use for this purpose the entropy
defined an appropriate way as
\begin{equation}\label{HE}
H(\hat{\cal E})=\int{\rm Tr}\,\hat{\cal E}(d\tilde\lambda) \ln
\frac{\hat{\cal E}(d\tilde\lambda)} {\nu(d\tilde\lambda)}\;.
\end{equation}

\noindent This expression defines entropy of the relative operator
measure $\hat{\cal E}$ with respect to the scalar measure $\nu$.
It generalizes usual definition $H^{\mu/\nu}=\int d\mu\ln
\frac{d\mu}{d\nu}$ of the relative entropy of the scalar measure
$\mu$ with respect to $\nu$. Measure $\nu$ in Eq. (\ref{HE}) is to
be chosen such a way that the optimal measure $\hat{\cal E}_0$
should be absolutely continuous with respect to $\nu$; to fit this
condition it is possible to apply the standard measure
$\nu(d\tilde\lambda)=d\tilde\lambda$\,.

Basing on these considerations, let us switch from the minimization
problem (\ref{min}) to the ``regularized'' problem
\begin{equation}\label{rmin}
{\cal R}_\varepsilon(\hat{\cal E})={\cal R}(\hat{\cal E})+\varepsilon
H(\hat{\cal E})=\min\limits_{\hat{\cal E}}\,
\end{equation}

\noindent where $\varepsilon>0$ is a small parameter similar to
the one used in Appendix A and analogous to a physical
temperature; ${\cal R} (\hat{\cal E})$ is an analogue of the
average energy of thermodynamics and $H(\hat{\cal E})$ is analogue
of the entropy. On the total, the problem (\ref{rmin}) is
analogous to the one of the statistical physics involved with the
deducing of the Gibbs distribution from the minimization of the
average energy at the condition of the fixed entropy value.
However, this problem differs from the thermodynamical one by the
operator-valued measure $\hat{\cal E}$ and the normalization
condition (\ref{cond}) (the second relation).  The solution
$\hat{\cal E}_\varepsilon$ of the problem (\ref{rmin}) on account
of the normalization condition (\ref{cond}) (the second relation)
is easy to get with use of the Lagrange multipliers in the form of
Gibbs-like distribution
\begin{equation}\label{Eeps}
\hat{\cal E}_\varepsilon(d\tilde\lambda)=\exp\left[\frac{\hat
F_\varepsilon-\hat R(\tilde\lambda)}{\varepsilon}\right] \nu(d
\tilde\lambda)\,.
\end{equation}

\noindent Here an undefined operator $\hat F$ is an operator analogue
of a $c$-number free energy $\varepsilon\Gamma(\varepsilon)$ in a
quantum Gibbs distribution $\hat\rho=\exp[\Gamma(\varepsilon)-\hat{\cal
H}/\varepsilon]$. It is calculated to fit the operator normalization
condition
\begin{equation}\label{opernorm}
\int\exp\left[\frac{\hat F-\hat R(\tilde\lambda)}{\varepsilon}\right]
\nu(d\tilde\lambda)=\hat I\,.
\end{equation}

If the solution $\hat F(\varepsilon)$ of this equation is acquired the
strictly optimal measure $\hat{\cal E}_0$ can be found by calculation
of the limit $\hat{\cal E}_0=\lim\limits_{\varepsilon\to0}\hat{\cal
E}_\varepsilon$. Corresponding to Eq. (\ref{Eeps}) entropy
$H=H(\hat{\cal E}_\varepsilon)$ and average risk ${\cal R}={\cal R}
(\hat{\cal E}_\varepsilon)$ can be expressed via $\hat F(\varepsilon)$
by formulas which are analogous to the thermodynamical ones:
$$
H={\rm Tr}\,\frac{d\hat F}{d\varepsilon},\quad {\cal R}={\rm Tr}\,\hat
F-\varepsilon H.
$$

\noindent (To prove this equation one can use the equality
$\displaystyle\frac{d}{d\varepsilon}\,{\rm Tr}\,\ln\int\hat{\cal
E}_\varepsilon(d \tilde\lambda)=0$, which is valid due to the
normalization condition.)

At this method the major complication is transferred to the operator
equation (\ref{opernorm}) for the free energy operator $\hat F$. If
operator $$\hat F_0=\varepsilon\ln\int\exp\left[-\frac{\hat R(\tilde
\lambda)}{\varepsilon}\right]\,\nu(d\tilde\lambda)$$

\noindent does repeatedly commute with all operators $\hat R(\tilde
\lambda)$, that is $$[[\hat R(\tilde\lambda),\hat F_0],\hat R(\tilde
\lambda)]=[[\hat R(\tilde\lambda),\hat F_0],\hat F_0]=0$$

\noindent for all $\tilde\lambda$, then it is a solution of Eq.
(\ref{opernorm}). It is easy to check substituting $\hat F_0$ into
Eq. (\ref{opernorm}) and using formula $\exp(\hat A+\hat
B)=\exp(\hat B/2)\exp(\hat A)\exp(\hat B/2)$ for repeatedly
commuting operators $\hat A$ and $\hat B$. Exactly to this
situation all the problems of Sec. \ref{section:lower},
\ref{section:reducing} can be reduced. Another possible
application of this method may be obtaining approximate solutions
with use of numeric calculations.

\section{Physical measurement procedure}\label{section:measurement}
Let us confine to the case of quantum decision function
\begin{equation}\label{EDec}
\hat{\cal
E}(d\tilde\lambda)=\hat{P}(\tilde\lambda)\,\frac{d\tilde\lambda}
{{\det}^{1/2}|2\pi C|}\;,
\end{equation}

\noindent where $\hat{P}(\tilde\lambda)$ are ``vacuum'' projectors for
operators $(\hat x-\tilde\lambda)^T(\hat x-\tilde\lambda)$ which are
quadratic on $\hat x$: $[\hat x,\hat x^T]=C\hat I$. As it follows from
Sec. \ref{section:lower}, \ref{section:reducing}, this function is
optimal for the case of the simple cost function and wide a'priori
distribution and for the case of quadratic cost function (when $\hat
x=\hat u$) and weak non-commutativity. Let us show that the directly
measured variables $\hat\Theta=\int\tilde\lambda\hat E(d\tilde\lambda)$
have the form
\begin{equation}\label{Theta}
\hat\Theta=\hat x+B\hat\xi,
\end{equation}

\noindent where operators $\hat\xi$ are copies of $\hat x$ in a
space $H'$ a copy of the space $H$, and $B$ fits the equation
\begin{equation}\label{BCB}
BCB^T=-C\,.
\end{equation}

\noindent The state $\hat\rho'$ of the additional system is
\begin{equation}\label{rpr}
\hat\rho'=\hat P'(0)
\end{equation}

\noindent where $\hat P'(0)$ is a copy of $\hat P(0)$.

The proof is given by check of the commutativity of the operators
(\ref{Theta}) and validness of the Eq.~(\ref{Edl}). For the
commutator $C_\Theta =[\hat\Theta,\hat\Theta^T]$ on account of Eq.
(\ref{BCB}) we have $C_\vartheta=C+BCB^T=0$. Then on acount o the
commutativity of $\hat\Theta$ we have
$$\hat E(d\tilde\lambda)=\delta(d\tilde\lambda,\hat\Theta)=
\delta(d\tilde\lambda,\hat x+B\hat\xi)=
(2\pi)^{-s}\int\exp[i\varkappa^T(\hat
x+B\hat\xi-\tilde\lambda)]d\varkappa d\tilde\lambda\,.$$

Substituting this expression into Eq.~(\ref{Edl}), where on
account of Eq.~(\ref{rpr}) is to be represented as Fourier
integral $\hat\rho'=
(2\pi)^{-s}\int\pi(i\tilde\varkappa)e^{i\tilde\varkappa\hat\xi}
d\tilde \varkappa$. Then we get
$${\rm Tr}_{H'}\hat E(d\tilde\lambda)\hat\rho'=d\tilde\lambda
(2\pi)^{-2s}\int\!\!\!\!\int\pi(i\tilde\varkappa) e^{i\varkappa^T(\hat
x-\tilde\lambda)}{\rm Tr}_{H'}e^{i\varkappa^TB\hat\xi+
i\tilde\varkappa^T\hat\xi}d\varkappa d\tilde\varkappa\,,
$$

\noindent where the commutativity of $\hat x$ and $\hat\xi$ and
possibility of confluence of the exponents of repeatedly commuting
operators under the ${\rm Tr}$ operation are taken into account.
Applying formula ${\rm Tr}\exp(i\mu^T\hat x)=(2\pi)^s{\det}^{-1/2}
|2\pi C|\delta(\mu)$ (which is easy to prove, e.g. with use of the
zero temperature method) and performing integration over
$\tilde\varkappa$ we get, on account of the symmetry of
$\pi(i\varkappa)$ and the relation $|\det B|=1$ valid due to Eq.
(\ref{BCB}), we get $${\rm Tr}_{H'}\hat
E(d\tilde\lambda)\hat\rho'=(2\pi)^{-s}{\det}^{-1/2}\int
\pi(iB^T\varkappa)e^{i\varkappa^TB(\hat x-\tilde\lambda)}
\frac{d\varkappa d\tilde\lambda}{{\det}^{1/2}|2\pi C|}=\hat
P(\tilde\lambda)\frac{d\tilde\lambda}{{\det}^{1/2}|2\pi C|}\;.
$$

\noindent Thus, the presented measurement procedure matches the
decision function (\ref{EDec}), Q. E. D.

\appendix

\section{Proof of the suggested properties of the family
$\hat\rho(\lambda)$ for an unlimitedly wide volume
$\tilde\Lambda\ni\lambda$}Now let us prove the above suggested
properties of the family
$\{\hat\rho(\lambda),\lambda\in\tilde\Lambda_i\}$ for an
unlimitedly expanding volumes $\tilde\Lambda_i\ni\lambda$,
$\delta\Lambda_i$. The statements are equal to the relations
\begin{eqnarray}
&\;\lim (\hat P_i\hat P_j-\delta_{ij}\hat P_i)=0,&\label{lim1}\\
&\!\lim \hat\rho(\lambda)\,\hat
P_i=\delta_{ij}\,\hat\rho(\lambda)&\mbox{ for
}\lambda\in\tilde\Lambda_j.\label{lim2}
\end{eqnarray}

\noindent The first one means that Eq. (\ref{PD}) defines projectors
onto mutually orthogonal subspaces $\Phi_i$, and the second one is the
condition of the non-trivial action of operators $\hat\rho(\lambda)$
only in the subspaces $\Phi_i$ corresponding to $\tilde\Lambda_i\ni
\lambda$.

The proof of the Eqs. (\ref{lim1}) and (\ref{lim2}) for $i\ne j$ is
capable to obtain, basing on the explicit expressions of the eigen
functions $\psi_0(\lambda)$, which may be used to represent the
projectors in the form $\hat P(\lambda)=\psi_0(\lambda)
\psi_0^+(\lambda)$ (this functions are frequently called ``coherent''
or ``Glauber'' states: for a two-dimensional, or --- in other terms
--- single-mode case $s=2$ the explicit form of $\psi_0(\lambda)$ is
given, for example, in Ref. \cite{Klauder}).

Vectors $\psi_0(\lambda)$ are asymptotically orthogonal to the
eigen functions $\psi_n(\lambda')$ of operators
$\hat\rho(\lambda')$ ($n=0,1,2,\dots$) if the distance
$\lambda'-\lambda$ with respect to the norm of commutator $C$
increases to infinity. This is exactly the case of unlimited
expansion of $\delta\Lambda_i$. In the case $i=j$ in (\ref{lim1})
and (\ref{lim2}) operators $\hat P_j$ should be represented with
the formula (\ref{PD}) and, on account of the above discussion,
the integration extended to the whole space $\Lambda$. Then $\hat
P_j$ will be changed to non-orthogonal expansion of unit in the
whole space $H$:
\begin{equation}\label{PH}
\hat I=\int\limits_{\Lambda}\hat P(\lambda)\,
\frac{d\lambda}{{\det}^{1/2}|2\pi C|}\,.
\end{equation}

To get the proof, it is most convenient to use the zero
temperature method. For this case it comes to representation of
projectors $\hat P(\lambda)$ in the form of the limit
\begin{equation}\label{Plim}
\hat P(\lambda)=\lim\limits_{\varepsilon\to0}
\hat\rho_\varepsilon(\lambda)
\end{equation}

\noindent of the density matrix
\begin{equation}\label{rhoe}
\hat\rho_\varepsilon=\exp\Biggl[\Gamma(\varepsilon)-\frac{(\hat{x}-
\lambda)^T(\hat{x}-\lambda)}{\varepsilon}\Biggr],
\end{equation}

\noindent which is a canonical Gibbs distribution for a
thermodynamic system with Hamiltonian $(\hat{x}-
\lambda)^T(\hat{x}-\lambda)$ and temperature $\varepsilon=kT$. Let
us substitute Eq. (\ref{Plim}) into Eq. (\ref{PH}) and replace
there $d\lambda$ by the gaussian measure $\mu(d\lambda)=
\exp(-\lambda^TK^{-1}_\mu\lambda/2)d\lambda$, which differs from
the probability one by absence of the normalizing multiplier
${\det}^{-1/2}(2\pi K_\mu)$ an falls to $d\lambda$ at the limit
$K_\mu\to\infty$. Then the modified integral is proportional to
marginal density matrix $\tilde{\rho}(\varepsilon, K_\mu)$ (with
the proportionality coefficient ${\det}^{1/2}(2\pi K_\mu)/
{\det}^{1/2}|2\pi C|$ absent) of the part $\hat x$ of the gaussian
collection $(\hat x,\lambda)$ with the joint quantum-classical
probability distribution
$\hat\rho_\varepsilon(\lambda){\det}^{-1/2}(2\pi K_\mu)
\mu(d\lambda)$. Density matrix $\tilde{\rho}(\varepsilon, K_\mu)$
is gaussian and its calculation algorithm reduces first, in
accordance to \cite{Stratonovich}, to calculation of the
correlation matrix $\tilde K=K_\varepsilon+K_\mu$, where in
accordance to Eq. (\ref{K}) $K_\varepsilon=(1/2)C\coth QC\approx
|C|/2$, then calculation from Eq. (\ref{K}) the corresponding
matrix
$$\tilde Q=C^{-1}\coth^{-1}(2\tilde{K}C^{-1})\approx(1/2)K_\mu^{-1},$$

\noindent and with use of Eq. (\ref{G}) --- the corresponding
normalization constant
$$e^{\tilde{\Gamma}}={\det}^{1/2}|2\sinh(\tilde{Q}C)|\approx
{\det}^{1/2}|2\pi C|/{\det}^{1/2}(2\pi K_\mu).$$

Finally, for the integral of interest we get
$$\begin{array}{l}\displaystyle\int\limits_{\Lambda}\hat P(\lambda)\,
\frac{d\lambda}{{\det}^{1/2}|2\pi
C|}=\lim\limits_{K_\mu\to\infty}\lim\limits_{\varepsilon\to0}
\frac{{\det}^{1/2}(2\pi K_\mu)}{{\det}^{1/2}|2\pi C|}
\int\limits_{\Lambda}\hat\rho_\varepsilon(\lambda)\frac{\mu(d\lambda)}
{{\det}^{1/2}(2\pi K_\mu)}=\\ \displaystyle
=\lim\limits_{K_\mu\to\infty}\lim\limits_{\varepsilon\to0}
\frac{{\det}^{1/2}(2\pi K_\mu)}{{\det}^{1/2}|2\pi C|}
\hat{\tilde\rho}(\varepsilon,K_\mu)=\lim\limits_{K_\mu\to\infty}
\exp\left(-\frac{1}{2}\hat x^TK^{-1}_\mu\hat x\right)=\hat I.
\end{array}$$

\noindent This proves Eq. (\ref{PH}) and completes the validation
of the above presented solution.

\begin{flushright}\small
Submitted 18 I 1972\normalsize
\end{flushright}

\end{document}